\documentclass[11pt]{article}
\usepackage{amsmath,amsthm}
\usepackage{amssymb,latexsym}
\usepackage[mathscr]{eucal}
\usepackage{setspace}
\usepackage{graphics}

\newcommand{\bee}{\begin{equation}}
\newcommand{\ene}{\end{equation}}

\newcommand{\al}{\alpha}

\begin{document}

\vspace*{.8in}

\begin{center}
{\bf Intermittency in the Enstrophy Cascade of \\Two-dimensional Fully-developed Turbulence: Universal Features}

\vspace{.5in}
Bhimsen K. Shivamoggi\footnote{Permanent Address: University of Central Florida, Orlando, Florida}\\
International Centre for Theoretical Physics\\
Trieste, Italy \end{center}

\vspace{.5in}

\doublespace

\noindent
{\bf Abstract}

Intermittency (externally induced) in the two-dimensional (2D) enstrophy cascade is shown to be able to maintain a {\it finite} enstrophy along with a vorticity conservation anomaly. Intermittency mechanisms of three-dimensional (3D) energy cascade and 2D enstrophy cascade in fully-developed turbulence (FDT) seem to have some universal features. The parabolic-profile approximation (PPA) for the singularity spectrum $f(\al )$ in multi-fractal model is used and extended to the appropriate {\it microscale} regimes to exhibit these features. The PPA is also shown to afford, unlike the generic multi-fractal model, an analytical calculation of probability distribution functions (PDF) of flow-variable gradients in these FDT cases and to describe intermittency corrections that complement those provided by the homogeneous fractal model.

\pagebreak

\section{Introduction}

Spatial intermittency is a common feature of fully developed turbulence (FDT) and implies that turbulence activity at small scales is not distributed uniformly throughout space. This leads to a violation of an assumption (Landau \cite{ll87}) in the Kolmogorov \cite{k41a} theory that the statistical quantities show no dependence in the inertial range $L\gg \ell\gg\eta$ on the large scale $L$ (where the external stirring mechanisms are influential) and the Kolmogorov microscale $\eta =(\nu^3/\varepsilon )^{1/4}$ (where the viscous effects become important),  $\varepsilon$ being the mean energy dissipation rate. Spatial intermittency effects can be very conveniently imagined to be related to the fractal aspects of the geometry of FDT (Mandelbrot \cite{m75}). The energy dissipation field may then be assumed to be a multi-fractal (Parisi and Frisch \cite{fp85}, Mandelbrot \cite{m89}). The latter idea has received experimental support (Meneveau and Sreenivasan \cite{ms91}).

In the multi-fractal model one stipulates that the fine-scale regime of FDT possesses a range of scaling exponents $\alpha \in I\equiv [\alpha_{\min} ,\alpha_{\max}]$. Each $\alpha \in I$ has the support set $\mathscr S(\alpha )\subset \mathbb R^3$ of fractal dimension (also called the singularity spectrum) $f(\alpha )$ such that, as $\ell \Rightarrow 0$, the velocity increment has the scaling behavior $\delta v (\ell )\sim\ell^\alpha$. The sets $\mathscr S (\alpha )$ are nested so that $\mathscr S (\alpha' )\subset \mathscr S (\alpha )$ for $\al ' <\al$.

Experimental data on three-dimensional (3D) FDT (Meneveau and Sreenivasan \cite{ms89}) suggested that the singularity spectrum function $f(\al )$ around its maximum may be expanded up to second order via the parabolic-profile approximation (PPA) \cite{ms89}\footnote{The PPA is equivalent to the log-normal model (Monin and Yaglom \cite{my75}).}$^,$\footnote{The PPA was shown to be a good fit for Lagrangian velocity statistics as well (Chevillard et al. [9]).} -

\begin{subequations}
\bee \label{eq1a}
f(\al )=f(\al_0 )+\frac12 f^{\prime\prime} (\al_0 )(\al -\al_0)^2
\ene
where,
\vspace*{-.1in}
\bee \label{eq1b}
f(\al_0)=3.
\ene
\end{subequations}

On the other hand, spatial intermittency in two-dimensional (2D) FDT has been a controversial issue. Indeed, even the whole theory of 2D FDT (Kraichnan \cite{rhk67} and Batchelor \cite{gkb69}) had, until recently, remained almost an academic exercise, notwithstanding its possible connections with atmospheric and oceanic large-scale flows. 2D FDT has now been produced to a close approximation in a variety of laboratory experiments (Couder \cite{yc84}, Kellay et al. \cite{kwg98}, Martin et al. \cite{mwgr98}, Rutgers \cite{mr98}, Rivera et al. \cite{rve98}, Vorobieff et al. \cite{vre99}, Rivera et al. \cite{rw02} and \cite{rdce03}). However, the Batchelor-Kraichnan theory for the enstrophy cascade in 2D FDT (with the energy spectrum $E(k)\sim k^{-3}$)\footnote{The Batchelor-Kraichnan energy spectrum was recently shown (Eyink \cite{Eyink}) to correspond to the inviscid limit of the Leray solutions of 2D Navier-Stokes equations. These solutions, however, have vorticity fields which exist only as {\it distributions}.} has essential differences with the Kolmogorov theory for the energy cascade in 3D FDT. The Batchelor-Kraichnan theory corresponds to flows with {\it infinite} enstrophy so there is no need for a nonlinearity-sustained cascade to transfer enstrophy across the inertial range to small scales to counter the dissipative action of viscosity there (Lopes Filho et. al. \cite{Filho}). Consequently, the usual cascade mechanism is not operational in the Batchelor-Kraichnan theory for the enstrophy regime.

Direct numerical simulations (DNS) of freely-decaying 2D FDT (McWilliams \cite{jcm84}, Benzi et al. \cite{bppsv86}, Brachet et al. \cite{bms86} and \cite{bmps88}, Kida \cite{sk85}, Ohkitani \cite{ko91}, Schneider and Farge \cite{sf05}) and forced-dissipative 2D FDT (Basdevant et al. \cite{blsb81}, Legras et al. \cite{lsb88} and Tsang et al. \cite{toag05})  showed intermittency caused by the presence of coherent structures. These structures inhibit the local inertial transfer of enstrophy via phase correlations across many length scales and produce energy spectra which are steeper than the Kraichnan-Batchelor spectrum $E(k)\sim k^{-3}$ for the enstrophy cascade.\footnote{There is, therefore, some speculation whether different universality classes exist in the enstrophy cascade depending on the particular initial conditions involved (\cite{bppsv86}, \cite{ko91} and \cite{lsb88}). On the other hand, the sensitivity of the 2D flow dynamics to the initial non-regularity of the flow raises questions about the whole universality concept in 2D FDT (Farge and Holschneider [32]).} 

The theroretical issue of intermittency in the 2D enstrophy cascade in view of the regular behavior of 2D Navier-Stokes solutions is a delicate one. In addition to coherent structures, another cause of intermittency in the 2D enstrophy cascade can be contamination from the 3D effects in any real flow situation. In the geophysical context, it can be caused by the Ekman drag (\cite{toag05}) which simulates the frictional planetary boundary layer (Pedlosky \cite{jp87}). The Ekman drag acts also as a sink at low wave numbers to take out the `condensate' from the inverse energy cascade. It is of interest to note that irrespective of the origin, an externally-induced intermittency is able to restore the usual nonlinearity-sustained cascade mechanism in the Batchelor-Kraichnan theory. Intermittency renders enstrophy of the flow {\it finite} so a nonlinearity - sustained cascade is very much needed to transfer enstrophy across the inertial range to small scales to counter the dissipative action of viscosity there. 

Paladin and Vulpiani \cite{pv87} pointed out that the intermittency in the enstrophy cascade may be described in terms of a multi-fractal probability measure for the vorticity gradient.\footnote{Indeed, experimental work of Jun and Wu [35] showed that large intermittency can be accounted for by the non-uniform distribution of saddle points in the flow (which are responsible for the energy transfer/dissipation (Daniel and Rutgers [36]).}. The multi-fractal characterization of the intermittent enstrophy dissipation field was given by Mizutani and Nakano \cite{mn89}, Benzi and Scardovelli \cite{bs95}, and Shivamoggi \cite{bks98}. In the intermittent case, the vorticity at very small length scales (where viscous effects are strong and nonlinearities are weak) behaves like a passive scalar (Weiss \cite{jw91}, Falkovich and Lebedev \cite{fl94} and Nam et al. \cite{noag00}) and is advected by the large-scale flow structures \cite{lsb88}. This leads to thin sheets with large vorticity gradients (Saffman \cite{pgs71}) and selective rapid decay of vorticity in these layers because such regions experience typically stronger viscous diffusion than other regions. Conversely, vorticity-gradient layers (or {\it divorticity} sheets) are more likely to occur near vortex nulls. (This is very akin to the vortex-sheet formation near velocity nulls and current-sheet formation near magnetic nulls in magnetohydrodynamics (Shivamoggi et al. \cite{Shiv1}-\cite{Shiv3})). On the other hand, the consequent statistical dependence between the vorticity and the vorticity gradient at a point may be expected in turn to lead to non-gaussian statistics. Indeed, probability density functions (PDF) of enstrophy flux was measured in a freely-decaying 2D FDT by Kellay et al. \cite{kbw95} which was found to be highly non-gaussian. On the otherhand, DNS (Herring and McWilliams \cite{hm85}, Borue \cite{vb93}, Maltrud and Vallis \cite{mv91}, Oetzel and Vallis \cite{ov97}), soap-film experiments (Gharib and Derango \cite{gd89}, Kellay et al. \cite{kwg95}, Martin et al. \cite{mwgr98}, Rutgers \cite{mr98}), electromagnetically driven flow experiments (Paret al. \cite{pjt99}) showed restoration of the Kraichnan-Batchelor scaling behavior when the coherent structures are suppressed.

The issue of whether intermittency (albeit externally-induced) in the 2D enstrophy cascade can maintain a {\it finite} enstrophy along with a vorticity conservation anomaly is of great interest - this is addressed in this paper. The possibility of common universal features (other than energy spectra) in the intermittency mechanisms of 2D and 3D FDT was raised by Dubrulle \cite{bd94} and is also a topic of great interest. We explore here universal features in the intermittency mechanisms of 3D energy cascade and 2D enstrophy cascade (when it is intermittent via external induction). The PPA (1) is used and extended to the appropriate {\it microscale} regimes for this purpose. We will also show that the PPA also affords,  unlike the generic multi-fractal model, an analytical calculation of PDF's of flow-variable gradients in these FDT cases.

\section{3D FDT: Energy Cascade}

\noindent
{\bf (i) Inertial Regime}

Let us briefly review and then extend the PPA applied to the inertial regime (Meneveau and Sreenivasan \cite{ms89}, Benzi and Biferale \cite{bb02}).

According to the multi-fractal model for the $p$th order velocity structure function (Parisi and Frisch \cite{fp85}), we have
\begin{subequations}
\bee \label{eq2a}
A_p \equiv \left< |\delta v|^p \right> \int \ell^{[p\al +3-f(\al )]} d\mu (\al ) \sim \ell^{\xi_p^{(1)}}
\ene
where,
\bee\label{eq2b}
\xi_p^{(1)} = \inf_{\al} \,  [p\al +3-f(\al )]
\ene
\end{subequations}
and this minimum occurs for $\al =\al^*$, which, according to the saddle-point method, is given by
\bee\label{eq3}
f'(\al_* )=p.
\ene

Writing \eqref{eq1a} and \eqref{eq1b} in the form -
\bee\label{eq4}
3-f(\al_*)=a(\al_* -\al_0)^2, \quad a>0
\ene
\eqref{eq3} yields,
\bee\label{eq5}
\al_*(p)=\al_0 - \frac{p}{2a} .
\ene

Using \eqref{eq4} and \eqref{eq5}, \eqref{eq2a} and \eqref{eq2b} become
\bee\label{eq6}
\xi_p^{(1)} =p \al_0 - \frac{p^2}{4a} .
\ene

The parameter $\al_0$ may now be determined (Benzi and Biferale \cite{bb02}) by using the exact 3D Navier-Stokes result (Kolmogorov \cite{ank41}) -
\bee\label{eq7}
\xi_3^{(1)} =1
\ene
which, on application to \eqref{eq6}, yields
\bee\label{eq8}
\al_0 =\frac13 +\frac{3}{4a} .
\ene

Using \eqref{eq8}, \eqref{eq6} becomes
\bee\label{eq9}
\xi_p^{(1)} =\frac{p}{3} +\frac{1}{4a} (3-p)p
\ene
while \eqref{eq5} becomes
\bee\label{eq10}
\al_* (p)=\frac13 +\frac{(3-2p)}{4a} .
\ene

Next, using \eqref{eq5} and \eqref{eq8}, \eqref{eq4} becomes
\begin{subequations}
\label{eq11a}
\bee
3-f(\al_*)=\frac{p^2}{4a} .
\ene
\end{subequations}

Upon interpreting $\varepsilon$ as the energy transfer rate (in the inertial regime, as Kraichnan \cite{rhk74} pointed out, the energy transfer rate rather than the energy dissipation rate is the dynamically important parameter), we have from (9),
\bee
\left< \varepsilon^2\right> \sim \ell^{-\frac{9}{2a}}.
\ene
Comparing (12) with the log-normal result \cite{my75} -
\bee
\left< \varepsilon^2\right> \sim\ell^{-\mu}
\ene
we have
\bee
\mu =\frac{9}{2a} .
\ene
Kolmogorov \cite{ank62} and Obukhov \cite{amo62} hypothesized that $\mu$ is a universal constant.

It is of interest to note the sense in which the PPA (1) mimics a multi-fractal in describing the intermittency aspects of FDT. Comparing \eqref{eq9} with the multi-fractal result (Meneveau and Sreenivasan \cite{ms91}) for the inertial regime -
\bee
\label{eq17}
\xi_p^{(1)} =\frac{p}{3} +\frac13 (3-p)(3-D_{p/3} )
\ene
we obtain
\begin{subequations}
\bee\label{eq18a}
D_{p/3} =3\left( 1-\frac{p}{4a} \right). 
\ene
\eqref{eq18a} implies
\bee\label{eq18b}
D_0=3 
\ene
\end{subequations}
which is also confirmed by (11a) that yields
\bee
f(\al_* (0))=3
\tag{11b}
\ene 
$f(\al_*(0))$ being the fractal dimension of the support of the measure, namely, $D_0$. It should be noted therefore that in the PPA the support of the measure is not a fractal; consequently, in the PPA the multi-fractality manifests itself via the way the {\it measure is distributed} rather than the {\it geometrical} properties like the support of the set.\footnote{A multi-fractal generalizes, as Mandelbrot \cite{m89} clarified, the notion of self-similarity from {\it sets} to {\it measures}.} In this sense the PPA is complementary to the homogeneous-fractal model (Frisch et al. \cite{fsn78}) in describing the intermittency aspects of FDT. This also implies that the failure to recognize the fractality of the support of the measure is apparently the cause of the well-known inability of PPA (See Castaing et al. \cite{cgh90}) to capture the quantitative aspects of intermittency adequately.

\vspace{.2in}

\noindent
{\bf (ii) Kolmogorov-Microscale Regime}

In order to determine universal features in the intermittency mechanisms of the various FDT cases, it is necessary to extend considerations to the Kolmogorov-{\it microscale} regime. Extension of the PPA to the latter regime has not been done. Let us now proceed to give this formulation.

On extending the multi-fractal scaling to the Kolmogorov-{\it microscale} $\eta_1$, where,
\bee\label{eq22}
\eta_1\sim\left(\frac{\nu^3}{\varepsilon}\right)^{1/4}
\ene
we have (Sreenivasan and Meneveau \cite{sm88} and Nelkin \cite{n90}) -
\bee \label{eq23}
B_p\equiv \left< \left| \frac{\partial v}{\partial x} \right|^p\right> \sim\int R_1^{-\frac{[p\al -p+3-f(\al )]}{1+\al}} d\mu (\al ),
\ene
where $R_1$ is the Reynolds number -
$$
R_1\sim\frac{( \varepsilon  L^4)^{1/3}}{\nu} \, .
$$

Saddle-point evaluation of the integral in (18) yields
\bee\label{eq24}
(1+\al_*)[p-f'(\al_*)]=p\al_*-p+3-f(\al_*).
\ene

Using \eqref{eq4}, \eqref{eq24} leads to
\bee\label{eq25}
a\al_*^2+2a\al_*+(2p-2a\al_0-a\al_0^2)=0
\ene
from which,
\bee \label{eq26}
\al_*(p)=-1\pm \sqrt{(\al_0+1)^2-\frac{2p}{a}} \; .
\ene

Imposing the condition -
\bee\label{eq27}
B_0\sim1
\ene
which, from \eqref{eq23}, implies
\bee\label{eq28}
\al_*(0)=\al_0
\ene
we see from \eqref{eq26} that the negative root needs to be discarded, and we obtain
\bee\label{eq29}
\al_*(p)=-1+\sqrt{(\al_0+1)^2-\frac{2p}{a}} \; .
\ene

On the other hand, using \eqref{eq4} and \eqref{eq24}, \eqref{eq23} yields
\bee\label{eq30}
B_p\sim R_1^{\gamma_p^{(1)}}
\ene
where,
\bee\label{eq31}
\gamma_p^{(1)}\equiv -[p+2a\{ \al_*(p)-\al_0\}].
\ene

In order to determine the parameter $\al_0$, the most pertinent framework for the Kolmogorov-{\it microscale} regime appears to be imposing the physical condition of {\it inviscid dissipation of energy} (IDE) (Kolmogorov \cite{ank41}). This implies
\bee\label{eq32}
\nu B_2\sim R_1^{\gamma_2^{(1)}-1} \sim \text{constant}
\ene
from which,
\bee\label{eq33}
\gamma_2^{(1)}-1=0.
\ene

Using \eqref{eq29} and \eqref{eq31}, \eqref{eq33} yields
\bee\label{eq34}
\al_0=\frac13 +\frac{3}{4a}
\ene
which is identical to \eqref{eq8} that was obtained by imposing the exact 3D Navier-Stokes result \eqref{eq7} in the inertial regime! This is  of course to be expected because the IDE is incorporated into the exact 3D Navier-Stokes result \eqref{eq7}.

Using \eqref{eq34}, \eqref{eq29} yields
\bee\label{eq35}
\al_*(p)=-1+\sqrt{\left(\frac43 +\frac{3}{4a}\right)^2 -\frac{2p}{a}}
\ene
while \eqref{eq31} then gives
\bee\label{eq36}
\gamma_p^{(1)}=\left[ p+2a\sqrt{\left(\frac{16a+9}{12a}\right)^2-\frac{2p}{a}}-\frac{16a+9}{6}\right]  \; .
\ene

For large $a$, \eqref{eq35} and \eqref{eq4} give the following asymptotic results -
\bee\label{eq37}
\al_*(p)=\frac13+(1-p) \frac{3}{4a}+O\left(\frac{1}{a^2}\right)
\ene
\bee\label{eq38}
3-f(\al_*)=\frac{9p^2}{16a}+O\left(\frac{1}{a^2}\right).
\ene
\eqref{eq37} and \eqref{eq38} show that the zero-intermittency limit corresponds to $a\Rightarrow\infty$, as before.

(32) shows (as (10) does) that
\bee
\al_*<\frac13 , \quad \forall p\geq 2,
\ene
implying of course the strengthening of the velocity-field singularities by intermittency in the Kolmogorov-{\it microscale} regime!

\vspace{.2in}
\noindent
{\bf (iii) Probability Distribution Function for the Velocity Gradient}

The multi-fractal model is known not to afford an analytical calculation of PDF of velocity gradient in intermittent FDT (Benzi et al. \cite{bbpvv91}). We now wish to show that the PPA is fruitful on this aspect. The physical principle underlying the calculation of the intermittency correction to the PDF of velocity gradient in the PPA turns out to be however the same as the one (namely, IDE) underlying the homogeneous-fractal model used in \cite{bbpvv91}.

Noting the scaling behavior of the velocity gradient (Frisch and She \cite{fs91}) -
\vspace*{-.1in}
\bee\label{eq100}
s\sim\frac{v}{\eta_1} \sim v_0^{\frac{2}{1+\al_*}} \nu^{\frac{\al_*-1}{1+\al_*}}
\ene
$v_0$ being the velocity increment characterizing large scales, and assuming $v_0$ to be gaussian distributed, i.e.,
\vspace*{-.1in}
\bee\label{eq101}
P(v_0)\sim e^{-\frac{v_0^2}{2\left< v_0^2\right>}}
\ene
we observe
\bee\label{eq102}
v_0^2\sim s^{(1+\al_*)} .
\ene
So, $\al_*(p)$ corresponds to $\al_*(\tilde p)$ where $\tilde p$ is the solution of
\bee\label{eq103}
\tilde p=1+\al_*(\tilde p).
\ene

Using \eqref{eq103}, and assuming $a$ to be large to simplify the calculations, we have from \eqref{eq37},
\bee\label{eq104}
\al_*(\tilde p)=\frac13 -\frac{1}{4a}+O\left( \frac{1}{a^2}\right) .
\ene

Using \eqref{eq104}, the PDF of the velocity gradient \cite{fs91} -
\bee\label{eq105}
P(s,\al_*(\tilde p))\sim\left(\frac{\nu}{|s|}\right)^{\frac{[1-\al_*(\tilde p)]}{2}} e^{- \left[ \frac{\nu^{\{ 1-\al_*(\tilde p)\}} |s|^{\{ 1+\al_*(\tilde p)\}}}{2\left< v_0^2\right>} \right]     }
\ene
becomes
\bee\label{eq106}
P(s,\al_*(\tilde p)) \sim \left(\frac{\nu}{|s|} \right)^{\left(\frac13 +\frac{1}{8a}\right)} e^{- \left[ \frac{\nu^{\left(\frac23 +\frac{1}{4a}\right)} |s|^{\left(\frac43 -\frac{1}{4a}\right)}}{2\left< v_0^2\right>} \right] } .
\ene

Incidentally, using \eqref{eq104}, \eqref{eq103} gives
\bee\label{eq107}
\tilde p=\frac43 -\frac{1}{4a} +O\left(\frac{1}{a^2}\right)
\ene
which is of course the exponent of $|s|$ in the argument of the exponential in \eqref{eq106} as to be expected from (37). Note the accentuation of the non-gaussianity of the PDF due to intermittency, as also indicated by the homogeneous-fractal model \cite{bbpvv91} which, as pointed out before, is however complementary to the PPA in describing the intermittency aspects of FDT.

\section{2D FDT: Enstrophy Cascade}

The evolution of vorticity in a 2D fluid flow is governed by 

\begin{subequations}
\begin{equation}
\frac{\partial \omega}{\partial t}+ ( \mathbf v \cdot \nabla)\omega=\nu\nabla^2 \omega
\end{equation}

\noindent
and is based on the competition between the viscous diffusion and the advection processes. 

Writing (43a) in the form

\begin{equation}
\frac{\partial \omega}{\partial t}+ \nabla\cdot (\mathbf{v}\omega)=\nu\nabla^2 \omega
\end{equation}
\end{subequations}

\noindent
we observe that vorticity is globally conserved in the absence of viscous diffusion, so the local maxima of vorticity can grow and possibilities of generation of non-Gaussian statistics for the vorticity PDF exist.

On the other hand, taking the gradient of (43a), we obtain

\begin{equation}
\frac{\partial}{\partial t} (\nabla\omega) + (\mathbf{v}\cdot\nabla)(\nabla\omega)=- \left( \nabla v_x \frac{\partial\omega}{\partial x} + \nabla v_y \frac{\partial\omega}{\partial y}\right)+\nu\nabla^2(\nabla\omega).
\end{equation}

(44) shows that the vorticity gradient is not conserved along the Lagrangian trajectories even in the absense of viscous diffusion, so the local maxima of the vorticity gradient field can grow. This would imply that the statistics of the vorticity gradient can also become non-Gaussian.

We therefore consider intermittency (albeit externally-induced) in the enstrophy cascade of 2D FDT and extend the PPA to formulate this.

{\bf (i) Inertial Regime}

The multi-fractal model for the $p$th order velocity structure function gives (Shivamoggi \cite{bks98})
\bee\label{eq49}
A_p\sim\ell^{\xi_p^{(2)}}
\ene
where,
\bee\label{eq50}
\xi_p^{(2)}=\inf_{\al} \, [p\al+2-f(\al )]
\ene
and this minimum $\al=\al_*$, according to the saddle-point method, corresponds to
\bee\label{eq51}
f'(\al_*)=p.
\ene

Assuming a PPA for the 2D case, we have -
\bee\label{eq52}
2-f(\al_*)=a(\al_*-\al_0)^2, \quad a>0
\ene
Using (48), (47) yields,
\bee\label{eq53}
\al_*(p)=\al_0-\frac{p}{2a} \, .
\ene

Using \eqref{eq52} and \eqref{eq53}, \eqref{eq50} becomes
\bee\label{eq54}
\xi_p^{(2)}=p\al_0-\frac{p^2}{4a} \,.
\ene

Using the {\it exact} result for the enstrophy cascade in 2D FDT (Eyink \cite{gle96} and Lindborg \cite{el99})\footnote{Equation (51) is also predicted by the multi-fractal model for the enstrophy cascade in 2D FDT (Shivamoggi \cite{bks98} and \cite{bks04}) and is in agreement with the rigorous inequalities established by Eyink \cite{gle95} and has been verified by DNS \cite{vb93}.} -
\bee
\xi_3^{(2)} =3
\ene
the parameter $\al_0$ may be determined  -
\bee\label{eq55}
\al_0=1+\frac{3}{4a}\,.
\ene

Using (55), (53) becomes
\bee\label{eq56}
\xi_p^{(2)}=p+\frac{1}{4a}(3-p)p
\ene
while (52) becomes
\bee\label{eq57}
\al_*(p)=1+\frac{(3-2p)}{4a}\,.
\ene

(54) shows that
\bee
\alpha_* (p)<1, \quad \forall p \geq 2
\ene
implying of course the strengthening of the vorticity-field singularities by intermittency! Indeed, (53) predicts an energy spectrum -

\bee
E(k)\sim k^{-3-\frac{1}{2a}}
\ene
which is steeper than $k^{-3}$, as required.

Observe that the intermittency corrections in (50) and (51) are identical to those for the energy cascade in 3D FDT, namely, (8) and (9)! 

On the other hand, (53) implies for the structure function of the vorticity field, the scaling behavior -
\bee
\left< | \delta\Omega |^p\right> \sim \ell^{\zeta_p^{(2)}}
\ene
where,
\bee
\zeta_p^{(2)} \equiv -\frac{3p}{4a} (p-3).
\ene
Comparing (57) with the multi-fractal result (Benzi and Scardovelli \cite{bs95}),\footnote{(58) is equivalent to \cite{bks98}
$$
\xi_p^{(2)} =p-\frac13 (p-3)(2-D_{p/3})
$$
which implies that the energy spectrum in the enstrophy cascade cannot be steeper than $k^{-11/3}$ (corresponding to $D_{p/3}=0$) in agreement with Sulem and Frisch \cite{sf75} and Pouquet \cite{ap78}. Further, the result $E(k)\sim k^{-11/3}$ agrees with the result of Gilbert \cite{adg88} which considered the dissipative structures to be line elements (with zero fractal dimension for their projections on the plane) - these lines are centers of accumulation of singularities associated with spiral vortex sheets in Gilbert's model.}
\bee
\zeta_p^{(2)}=-\frac13 (p-3)(2-D_{p/3} )
\ene
we obtain
\begin{subequations}
\bee
D_{p/3} =2-\frac{3p}{4a} \, .
\ene
Observe that the intermittency correction in (60a) is again identical to that for the energy cascade in 3D FDT, namely, (16a)! Incidentally, the necessity of negative H\"older singularities of the vorticity field to produce intermittency in the enstrophy cascade and to preclude enstrophy conservation in the inviscid limit was noted by Eyink \cite{gle96}.

On the other hand, (60a) implies

\bee
D_0 = 2
\ene
\end{subequations}
which signifies that in the PPA the support of the enstrophy-dissipation field in 2D FDT is not a fractal. This has been confirmed for 2D FDT however by the DNS [30].

These results appear to signify universal features in the intermittency mechanisms of 3D energy cascade and 2D enstrophy cascade.

\noindent
{\bf (ii) Kraichnan Microscale Regime}

On extending the multi-fractal scaling to the Kraichnan {\it microscale} $\eta_2$ (Shivamoggi \cite{bks90}), where,
\bee\label{eq58}
\eta_2\sim\left(\frac{\nu^3}{\tau}\right)^{1/6}
\ene
$\tau$ being the mean enstrophy dissipation rate, we have  \cite{bks98} -
\bee\label{eq59}
C_p\equiv\left<\left|\frac{\partial^2v}{\partial x^2}\right|^p\right>\sim\int R_2^{-\frac{[p\al -2p+2-f(\al )]}{1+\al}} d\mu (\al )
\ene
where $R_2$ is the Reynolds number for the 2D FDT -
$$
R_2 \sim\frac{(\tau L^6)^{1/3}}{\nu} \,.
$$

Saddle-point evaluation of the integral in \eqref{eq59} yields
\bee\label{eq60}
(1+\al_*)[p-f'(\al_*)]=p\al_*-2p+2-f(\al_*).
\ene

Using \eqref{eq52}, \eqref{eq60} leads to
\bee\label{eq61}
a\al_*^2+2a\al_*+(3p-2a\al_0-a\al_0^2)=0
\ene
from which,
\bee\label{eq62}
\al_*(p)=-1\pm\sqrt{(\al_0+1)^2-\frac{3p}{a}} \;.
\ene

Imposing the condition -
\bee\label{eq63}
C_0\sim 1
\ene
on (62), we have,
\bee\label{eq64}
\al_*(0)=\al_0.
\ene
Using (67), we see that the negative root in (65) is to be discarded, and we obtain
\bee\label{eq65}
\al_*(p)=-1+\sqrt{(\al_0+1)^2-\frac{3p}{a}}\;.
\ene

On the other hand, using (48) and (63), (67) yields
\bee\label{eq66}
C_p\sim R_2^{\gamma_p^{(2)}}
\ene
where,
\bee\label{eq67}
\gamma_p^{(2)}\equiv -[p+2a\{\al_*(p)-\al_0\}].
\ene

In order to determine the parameter $\al_0$, the most pertinent framework for the Kraichnan-{\it microscale} regime in the 2D enstrophy cascade appears to be imposing the physical condition of {\it inviscid dissipation of enstrophy} (ID$\hat{\text{E}}$) \cite{bks98}.\footnote{Indeed, Polyakov [75] argued that the existence of the enstrophy cascade is predicated on the enstrophy conservation law anomaly, while recent numerical computations (Dmitruk and Montgomery \cite{Dmit}) showed that the tendency, if any, of enstrophy dissipation to go to zero in the inviscid limit is a very weak one.} This implies
\bee\label{eq68}
\nu C_2\sim R_2^{\gamma_2^{(2)}-1}\sim\text{const}
\ene
from which,
\bee\label{eq69}
\gamma_2^{(2)}-1=0.
\ene

Using (68) and (70), (72) yields
\bee\label{eq70}
\al_0=1+\frac{3}{4a}
\ene
which is identical to (52) that was obtained by imposing the  2D multi-fractal model result (51) in the inertial regime! This appears to indicate that the ID$\hat{\text{E}}$ has been incorporated into the result (51) in a manner similar to the case with the Kolmogorov exact result in the 3D incompressible case. Thus, intermittency (albeit externally-induced) can maintain a {\it finite} enstrophy along with a vorticity conservation anomaly without contradicting the rigorous result in [20] and [21].

Using (73), (68) yields
\bee\label{eq71}
\al_*(p)=-1+\sqrt{\left( 2+\frac{3}{4a}\right)^2-\frac{3p}{a}}
\ene
while (70) then gives
\bee\label{eq72}
\gamma_p^{(2)} =-\left[ p+2a\sqrt{\left(\frac{3+8a}{4a}\right)^2-\frac{3p}{a}} -\frac{3+8a}{2}\right] .
\ene

The intermittency corrections $\Delta \alpha_* (p)$ to the scaling exponent $\alpha_*(p)$ for the 3D (given by (30)) and the 2D (given by (74)) are sketched in Figures 1a and 1b for different values of the intermittency parameter $a$. Observe that the velocity-field singularities are strengthened in the {\it microscale} regimes by the intermittency effects. Intermittency corrections $\Delta\alpha_* (p)$ for 2D cases are smaller than those for 3D cases. Further, in the weak-intermittency limit ($a$ large), the intermittency corrections $\Delta \alpha_* (p)$ for the 3D and 2D cases are almost identical.

For large $a$, (74) and (48) give the following asymptotic results -
\bee\label{eq73}
\al_*(p)=1+(1-p)\frac{3}{4a}+O\left(\frac{1}{a^2}\right)
\ene
\bee\label{eq74}
2-f(\al_*)=\frac{9p^2}{16a}+O\left(\frac{1}{a^2}\right).
\ene
Observe that the intermittency corrections in (76) and (77) for the 2D enstrophy cascade {\it microscale} regime are identical to those in (32) and (33) for the 3D energy cascade {\it microscale} regime! This appears to confirm further the universal features in the intermittency mechanisms of 3D energy cascade and 2D enstrophy cascade pointed out earlier.

\vspace{.2in}
\noindent
{\bf (iii) Probability Density Function for the Vorticity-Gradient}

Noting the scaling behavior of the vorticity gradient (Shivamoggi \cite{bks98}) -
\bee\label{eq108}
r\sim\frac{v}{\eta_2^2} \sim v_0^{\frac{3}{1+\al_*}} \nu^{\frac{\al_*-2}{1+\al_*}}
\ene
and assuming \eqref{eq101} again, we observe
\bee\label{eq109}
v_0^2\sim r^{\frac23 (1+\al_*)} .
\ene
So, $\al_*(p)$ corresponds to $\al_*(\tilde p)$ where $\tilde p$ is now the solution of
\bee\label{eq110}
\tilde p=\frac23 [1+\al_*(\tilde p)].
\ene

Using (80), and assuming again $a$ to be large, we have from (76),
\bee\label{eq111}
\al_*(\tilde p)= 1-\frac{1}{4a} +O\left( \frac{1}{a^2}\right).
\ene

Using (81), the PDF of the vorticity gradient \cite{bks98} -
\bee\label{eq112}
P(r,\al_*(\tilde p))\sim\left(\frac{\nu}{|r|}\right)^{\frac{[2-\al_*(\tilde p)]}{3}} e^{-\left[ \frac{\nu^{\frac23 \{ 2-\al_*(\tilde p)\}} |r|^{\frac23 \{ 1+\al_*(\tilde p)\}}}{2\left< v_0^2\right>} \right] }
\ene
becomes
\bee\label{eq113}
P(r,\al_*(\tilde p))\sim\left(\frac{\nu}{|r|}\right)^{\frac13\left( 1+\frac{1}{4a}\right)} e^{- \left[ \frac{\nu^{\frac23 \left( 1+\frac{1}{4a}\right)} |r|^{\frac23\left( 2-\frac{1}{4a}\right)}}{2\left< v_0^2\right> } \right]} .
\ene

Using (81), (80) gives
\bee\label{eq114}
\tilde p =\frac23\left( 2-\frac{1}{4a}\right) +O\left(\frac{1}{a^2}\right)
\ene
which is again the exponent of $|r|$ in the argument of the exponential in (83), as to be expected from (79). Note again the accentuation of the non-gaussianity of the PDF due to intermittency, as also indicated by the homogeneous-fractal model \cite{bks90} which is however complementary to the PPA in describing the intermittency aspects of FDT.

\section{Conclusions}

The theoretical issue of intermittency in the 2D enstrophy cascade in view of the regular behavior of 2D Navier-Stokes solutions is a delicate one. It is of interest to note that, irrespective of the origin, an externally-induced intermittency is able to restore the usual nonlinearity-sustained cascade mechanism in the Batchelor-Kraichnan theory. Intermittency renders enstrophy of the flow {\it finite} so a nonlinearity sustained cascade is very much needed to transfer enstrophy across the inertial range to small scales to counter the dissipative action of viscosity there. Further, as we have shown in Section 3 (ii), intermittency can maintain a {\it finite} enstrophy along with a vorticity conservation anomaly without contradicting the rigorous result in [20] and [21].

Intermittency mechanisms of 3D energy cascade and 2D enstrophy cascade appear to have certain universal features notwithstanding very different physics underlying these FDT cases. This probably results, following Kadanoff's \cite{lk90} speculation, because of consideration of the limit $R \Rightarrow\infty$, which happens to be the critical point for FDT. Indeed, universal features for very diverse FDT cases become apparent at the critical point (Shivamoggi \cite{bks05}). On the other hand, via (14) this appears to support the universality of the log-normal exponent $\mu$ hypothesized by Kolmogorov \cite{ank62} and Obukhov \cite{amo62}. It remains to be pointed out, however, though the PPA appears to have the capacity to provide considerable insight into the qualitative aspects of intermittent FDT quantitative aspects have been a different story for the 3D FDT problem (see Castaing et al. \cite{cgh90}). This is probably traceable to the failure of PPA to recognize the fractality of the support of the measure, as shown in this paper.

\vspace{.2in}
\noindent
\large{\bf Acknowlegements}
\normalsize

This work was carried out during the author's visit to the  International Centre for Theoretical Physics, Trieste, Italy. The author is thankful to Professor Katepalli Sreenivasan for his valuable remarks and discussions.

\singlespace

\newpage

\begin{center}
\scalebox{.5}{\includegraphics{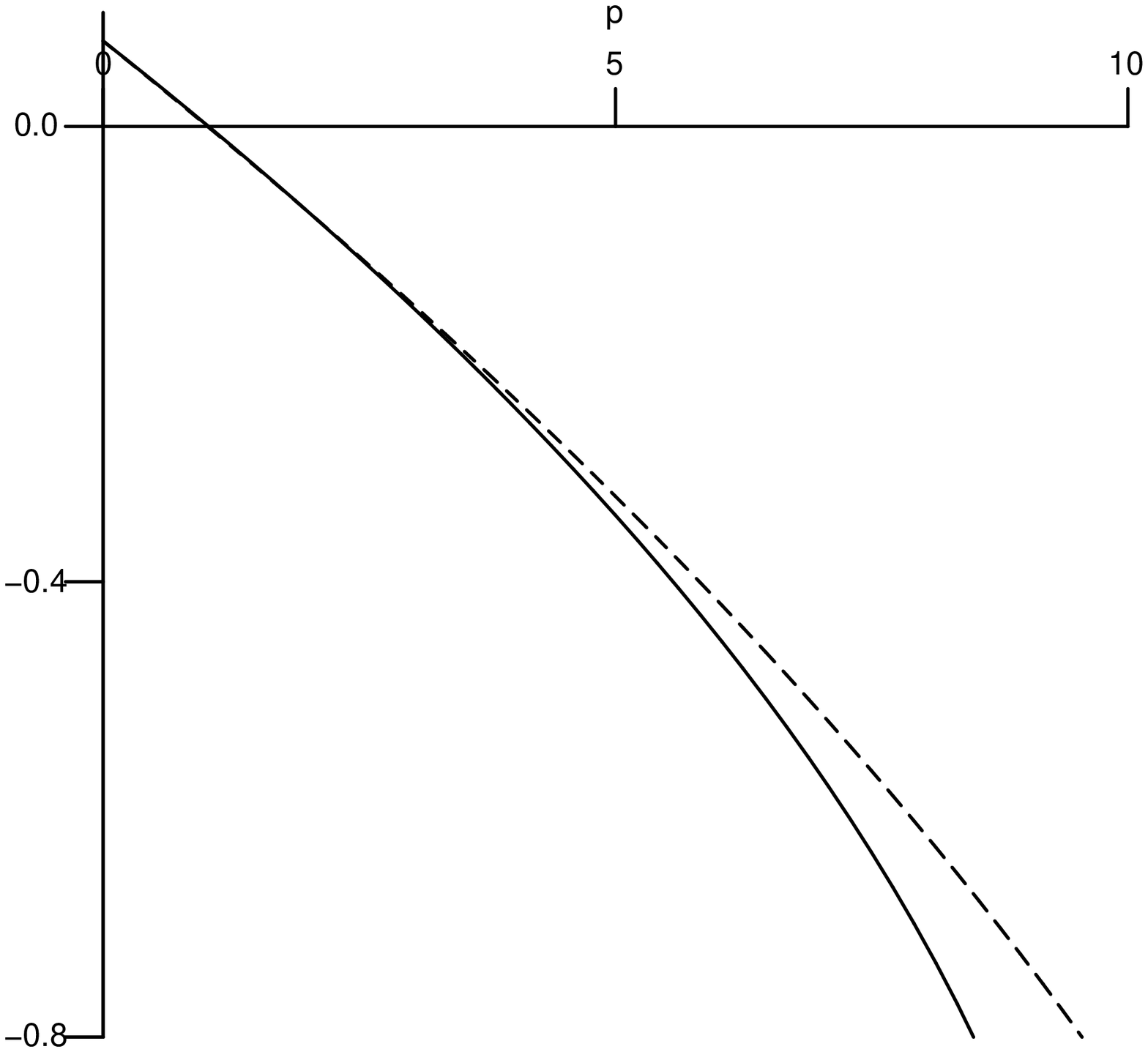}}
\end{center}

\begin{singlespace}

\vspace*{-.3in}

\begin{center} Figure 1a. Intermittency correction $\Delta\alpha_*(p)$ vs. $p$ for the intermittency parameter \\
$a=10$ (--- 3D,  - - -  2D). \hspace*{2.2in} \end{center}

\begin{center}
\scalebox{.5}{\includegraphics{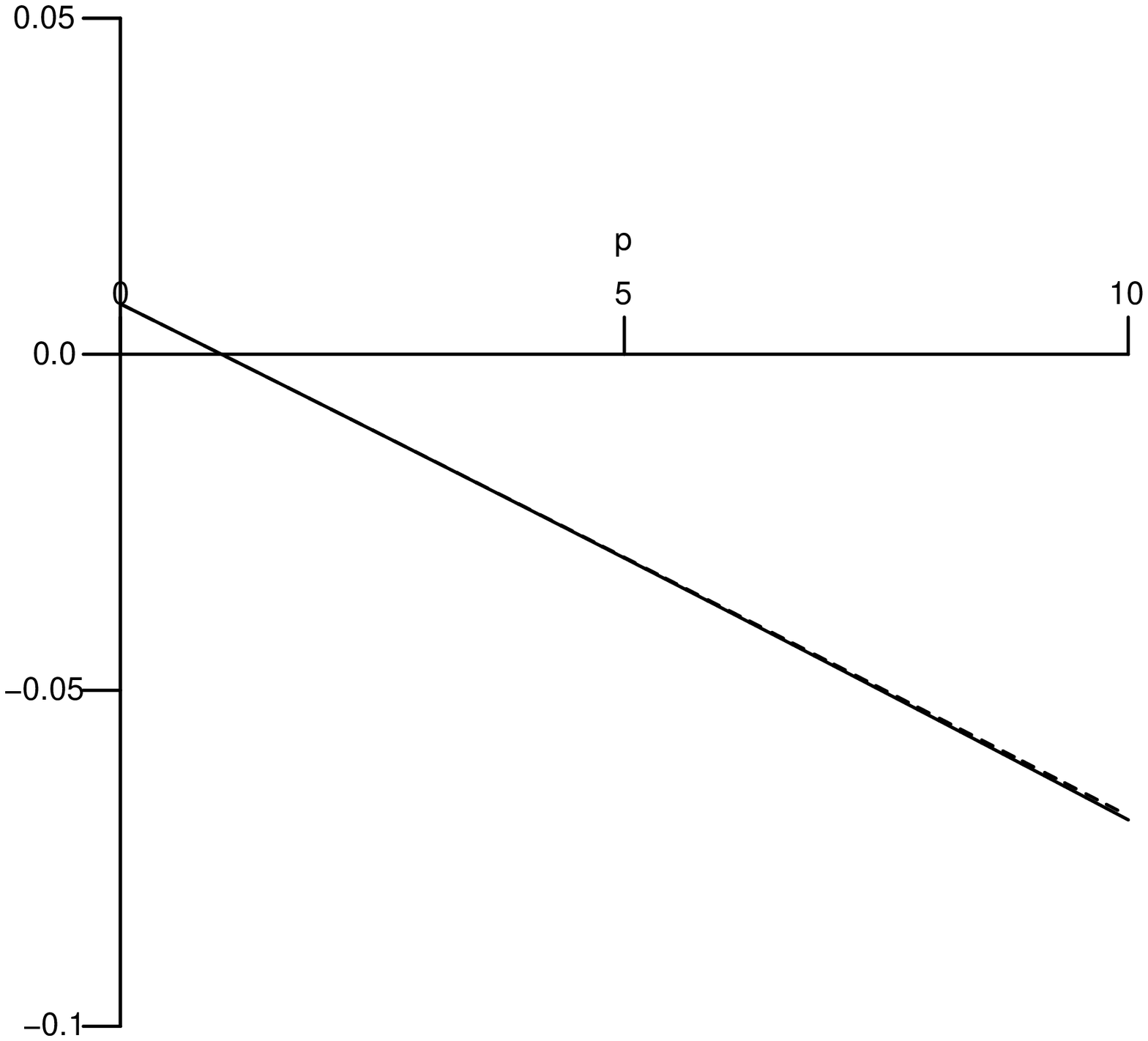}}
\end{center}

\vspace*{-.3in}

\begin{center} Figure 1b. Intermittency correction $\Delta\alpha_*(p)$ vs. $p$ for the intermittency parameter \\
$a=100$ (--- 3D,  - - -  2D). \hspace*{2.2in} \end{center}

\end{singlespace}

\end{document}